\documentclass[12pt]{article}
\usepackage{graphicx}
\usepackage{subfigure}
\usepackage{amssymb}
\usepackage{amsfonts}
\usepackage{latexsym}

\usepackage[usenames]{color}
\usepackage{latexsym}
\usepackage{epstopdf} %%%
\usepackage[normalem]{ulem}
\newcommand{\beq}{\begin{equation}}
\newcommand{\eeq}{\end{equation}}
\newcommand{\be}{\begin{equation}}
\newcommand{\ee}{\end{equation}}
\newcommand{\bea}{\begin{eqnarray}}
\newcommand{\eea}{\end{eqnarray}}

\makeatletter

\@addtoreset{equation}{section}
\makeatother
\def\href#1#2{#2}
\usepackage{amsmath}	% required for `\align' (yatex added)
\begin{document}

\baselineskip=15.5pt
\pagestyle{plain}
\setcounter{page}{1}

\begin{titlepage}
\begin{flushleft}
      % \hfill                      {\tt hep-th/1102.****}\\
       \hfill                       FIT HE - 19-01 \\
       \hfill                       %KYUSHU-HET 132 \\
\end{flushleft}
%\vspace*{3mm}

\begin{center}
  {\huge Color Superconductivity\\   
   \vspace*{2mm}
in a Holographic Model %SYM Theory %in hyperbolic space-time 
\vspace*{2mm}
}
\end{center}
%\vspace{5mm}

\begin{center}

%\vspace*{2mm}
\vspace*{5mm}
{\large ${}^{\dagger}$Kazuo Ghoroku\footnote[1]{\tt gouroku@fit.ac.jp},
${}^{\dagger}$Kouji Kashiwa\footnote[2]{\tt kashiwa@fit.ac.jp},
%${}^{\ddagger}$
Yoshimasa Nakano\footnote[3]{\tt ynakano@kyudai.jp},\\
${}^{\S}$Motoi Tachibana\footnote[4]{\tt motoi@cc.saga-u.ac.jp}
and ${}^{\ddagger}$Fumihiko Toyoda\footnote[5]{\tt ftoyoda@fuk.kindai.ac.jp}
}\\

%\vspace*{5mm}
\vspace*{2mm}
{${}^{\dagger}$Fukuoka Institute of Technology, Wajiro, 
Fukuoka 811-0295, Japan\\}
\vspace*{2mm}
{${}^{\S}$Department of Physics, Saga University, Saga 840-8502, Japan\\}
\vspace*{2mm}
%\large 
{${}^{\ddagger}$Faculty of Humanity-Oriented Science and
Engineering, Kinki University,\\ Iizuka 820-8555, Japan}
\vspace*{3mm}
\end{center}

\begin{center}
{\large Abstract}
\end{center}

A holographic bottom-up model used in studying the superconducting system 
is applied to search for the color superconducting phase of supersymmetric
Yang-Mills theory {with quarks}. We apply the probe analysis of this model to the supersymmetric Yang-Mills
theory in both the confinement and deconfinement phases.
In this analysis, we find the color superconductivity in both phases
when the baryon chemical potential exceeds a certain critical value.
This result implies that, above the critical chemical potential, 
a color non-singlet diquark operator, namely the Cooper pair,
has its vacuum expectation value even in the confinement phase.
In order to improve this peculiar situation, we proceed the analysis 
by taking account of the full back-reaction from the probe.
As a result, the color superconducting phase, which is observed in the probe approximation, 
disappears in both the confinement and deconfinement phases
when parameters of the theory are set within their reasonable values.

\noindent

%\vfill
\begin{flushleft}

\end{flushleft}
\end{titlepage}
%\newpage

\vspace{1cm}
%%%%%%%%%%%%%%%%%%%%%%%%%%%%%%%%%%%%%%%%%%%%%%%
\section{Introduction}

The color superconducting (CSC) phase has been expected in QCD
at finite baryon chemical potential,
but it is difficult to show it (see for example the review \cite{Alford:2007xm}). % analytically.
The numerical simulation is also difficult due to the complex
Euclidean action problem known as the sign problem. % \kg{(see for example \cite{deForcrand:2010ys})}.
%method through the lattice theorythrough
%with quarks in terms
On the one hand, the gauge/gravity duality\cite{ads1,ads2,ads3}
has been used as a powerful method to clarify many properties related to QCD
in the strong coupling region. 
As for the CSC, however, the holographic method is not clear
to be useful.
The reason is that we need to introduce a color non-singlet scalar field,
which provides us the vacuum expectation value (VEV)
of the color non-singlet operator in the CSC phase. 
In the top-down holographic approaches, the scalar field introduced
both in the bulk and the probe branes should be a color singlet. 

On the other hand, from the viewpoint of bottom-up approach,
a holographic model for the superconducting phase %of some non-color charge 
has been studied by introducing a $U(1)$ gauge field
and a charged scalar field in the bulk\cite{Gub,Hart,Hartnoll:2008kx,Nishi}.
The important point in applying them as the holographic CSC model %in those models
is that the charge of the $U(1)$ gauge field is regarded
as the baryon number\cite{Fadafan,Basu}. {Namely, the $U(1)$ gauge field is dual to 
the baryon number current, density and chemical potential for the quarks of $SU(N_c)$ 
Yang -Mills theory. And, the charged scalar is dual to a composite field operator with a finite quark number.
In the model, they are not introduced through the probe D-branes but given by the bulk action. Although it is not known how this
model is lifted up to the ten dimensional superstring theory, we call the model considered here 
the holographic supersymmetric Yang-Mills (SYM) theory by respecting the basic gauge theory dual to the gravity in the AdS$_5$.
}

In Ref.~\cite{Fadafan}, the matter part,
the system consisting of a $U(1)$ gauge field and a charged scalar field,
is treated as the probe as in Ref.~\cite{Hart}.
In other words, any back-reaction from the matter part %$L_\mathrm{CSC}$ 
to the bulk configuration is neglected\footnote{
This approximation would be justified for the case of large charge
of the scalar, however, we do not impose this restriction here.}.
%In this case,
Then the authors of Ref.~\cite{Fadafan} performed the analysis
in the high temperature deconfinement phase of the SYM % supersymmetric Yang-Mills
theory, which is dual to the five dimensional AdS-Schwarzschild background, 
and found the CSC phase above a certain chemical potential.

%In \cite{Basu,Fadafan}, this model
%, which is the d+1 dimensional gravity with $U(1)$ gauge field and a charged scalar,
%has been used to study the color superconducting phase %d(=5 and 4)-dimensional 
%of the $SU(N_{\rm c})$ SYM theory. 

In Ref.~\cite{Basu}, on the other hand, the back-reaction from the matter part to the bulk
is fully taken into account.
Due to the back-reaction from the $U(1)$ gauge field,
the high temperature deconfinement phase is described
by the Reissner-Nordstrom charged black hole in the gravity side.
%, which is known as a solution of the Maxwell-Einstein equation. 
%%% However, a color neutral, and baryon number charge $q=0$, scalar field
% This line is changed into the following line.
In Ref.~\cite{Basu}, however, a color neutral scalar field
%with no baryon number
is chosen %as a dual operator corresponding 
to study the color superconductor.
Furthermore, the conformal dimension of the scalar
is supposed to be %different from 
smaller than that of the diquark operator.
As a result of these special setting,
at very small temperature, a phase transition to the CSC phase
has been observed.

{In this paper,} taking the idea and technique of Refs.~\cite{Hart,Fadafan,Basu} into account, 
%following the viewpoint of Refs.\cite{Hart,Fadafan,Basu}, 
we proceed the analysis of the bottom-up model %(\ref{bottom-up}) to
in order to get more profound knowledge about the CSC phase in the SYM theory.
At first, we apply a simple probe analysis, in which the matter part is treated 
%$L_\mathrm{CSC}$ 
as a probe, to the background for both the confinement and deconfinement phases. 
In this analysis, the CSC phase is confirmed by the non-vanishing VEV
of a color non-singlet diquark operator, namely the Cooper pair. 
%In a simple probe approximation \kg{used in this paper, we just}
As a result, we find CSC phase %this phase transition also 
in both the deconfinement and confinement backgrounds.
%\footnote{
%See Fig.~\ref{Phase-diagram-2-1}.}. 
At a glance, %the phase transition in the confining phase 
it seems strange that the VEV of a color non-singlet operator exists
in the confinement phase.
%bulk configuration corresponds to the confinement phase.
%One may wonder whether is it meaningful to put on the probe
%on the confinement background.
This may indicate the breakdown of the simple probe approximation.
The probe approximation is available for large scalar charge, which
is supposed as $q=2/N_c$ in this paper, and the holographic approach is useful for large $N_c$.
Then the %probe approximation should be evaded. %
two methods are not compatible. 
We therefore cannot trust the results obtained from the probe approximation.

%The confinement and deconfinement phases used in the probe approximation are given
%by the Einstein-Hilbert %different solutions of the bulk action which is independent of the chemical potential.
%Instead,  %
In order to improve the probe approximation, %other words, 
it is natural to consider the vacuum solutions 
%phase diagram,
which are given by solving the Einstein-Maxwell equation of the
system with the Einstein-Hilbert action and the $U(1)$ gauge field part.
%the bulk solution, 
In this case, the phase diagram in $\mu$-$T$ plane should be modified to the chemical potential
dependent form.
\footnote{
In Ref.~\cite{Nishi}, in a context of the R-charge superconductor,
an analysis has been done based on the similar background configuration.}
% $L_\mathrm{CSC}$. 
%for the model (\ref{bottom-up}) by including the effect comming from
%the chemical potential in the probe part.
This modification is equivalently obtained
%This point can be improved 
by taking account of the full back-reaction
from the probe action as shown in Ref.~\cite{Basu}.
Thus, as the next step, we proceed the analysis based on this modification. %method in Ref.~\cite{Basu}. 
%Notice, however, the mass and the $U(1)$ charge for the scalar field
%are set differently from  Ref.~\cite{Basu}.

\vspace{.3cm}
In the next section, we set up our holographic model,
and make a probe analysis to find the CSC phase in SYM theory. 
In the resultant phase diagram
obtained by the probe approximation, we find a result which is unacceptable from the viewpoint of QCD.
%The probe part consists of the $U(1)$ gauge field and a charged scalar field. 
In Sec.~\ref{sec:backreaction}, we continue the analysis by taking
account of the back-reaction to improve the probe approximation,
and we search the CSC phase in the improved background.
%studied as a probe action in the different improved background. 
Summary and discussions are given in the final section.
%In  Appendix~\ref{sec:originOpsi}, the effective mass of the scalar field is explained in relation to the phase transition. 

%-----------------------------------------------------------------------

%\newpage
\section{A bottom-up model and a probe approach}

We consider a bottom-up %the following % the probe method for finding 
holographic dual for the SYM theory.  %\footnote{\kg{We don't know how it is related to the ten dimensional string theory.}}
It is given by the following 
%(d+1)-dimensional 
gravitational theory\cite{Gub,Hart}.
%S_6 &=& \int d^{d+1} x \sqrt{-g} \left\{ {\cal R} + {20 \over L^2} + L_{CSC}  \right\},
\bea\label{bottom-up}
S &=& \int d^{d+1} x \sqrt{-g}\,\mathcal{L}  \,,  \\
 \mathcal{L}&=&\mathcal{L}_\mathrm{Gravity}+\mathcal{L}_\mathrm{CSC} \label{action}\,,   \label{action-1} \\
     \mathcal{L}_\mathrm{Gravity}&=&  {\cal R} + {d(d-1) \over L^2}\,, \label{bulk-L} \\
     \mathcal{L}_\mathrm{CSC} &=& - {1 \over 4} F^2 - |D_{_\mu} \psi|^2 - m^2 |\psi|^2\, , \label{probe-L} \\
   F_{\mu\nu}&=&\partial_\mu A_\nu-\partial_\nu A_\mu\,,\quad
D_{\mu} \psi = (\partial_{\mu}-iqA_{\mu})\psi\, .
\eea
%This is a simple bottom-up model proposed by . 
It describes $d+1$ dimensional gravity coupled to a $U(1)$ gauge field, $A_\mu$, %$F_{\mu\nu}=\partial_{\mu} A_{\nu}-\partial_{\nu} A_{\mu}$,
and a charged scalar field, $\psi$. The charge $q$ denotes the baryon number 
of the scalar $\psi$, {which is considered to be dual to the diquark operator in this paper,}
and for the moment it is chosen that $q=2/3$. \footnote{
In the model, it should be taken as $q=2/N_c$. In this section, we suppose that we are considering
the dual theory as QCD with $SU(3)$ gauge group.}
The mass $m$ is given to reproduce the correct conformal dimension of the diquark
operator dual to the scalar field $\psi$. 
Here we put $1/2\kappa_6^2 =1$, and consider the case of $d=5$ hereafter.

The above holographic model, previously, has been considered to be dual to the superconductor of the electric charge \cite{Gub,Hart}
and of the R-charge \cite{Nishi}. And it is recently extended to a theory dual to the color superconductor in \cite{Fadafan,Basu}.
However, it is unknown
how this theory is dual to the SYM theory and can be related to the ten dimensional string theory. 
In this paper, we proceed the analysis of this model by supposing
that this is dual to the SYM theory to study the existence of the CSC phase.

\vspace{1cm}
\noindent{\bf Bulk and Probe}

Here, 
$\mathcal{L}_\mathrm{CSC}$ is considered as the probe to see the condensation
of the colored operator which is expressed in terms of the scalar field $\psi$. 
Therefore, the vacuum of the dual SYM theory is given by the solution of the Einstein equation of the action,
\beq\label{Gravity}
S_\mathrm{Gravity} = \int d^{d+1} x \sqrt{-g} \left\{ {\cal R} + {20 \over L^2}\right\}\, ,
\eeq
where $\mathcal{L}_\mathrm{CSC}$ is neglected.
Hereafter, we put $L=1$.
After giving the vacuum of the SYM theory by solving the above action, 
the equations of motion of
the probe action  $\mathcal{L}_\mathrm{CSC}$ are solved without changing the background configuration given by $S_\mathrm{Gravity}$. 

%At first we solve the gravitational equation of motion without $\mathcal{L}_\mathrm{CSC}$. The solution represents the vacuum state of the
%SYM theory when its asymptotic form shows the AdS geometry.

In the present case, we could find two solutions of the Einstein equation of (\ref{Gravity}). 
They represent a low temperature confinement phase
and a high temperature deconfinement one.
For each phase, we study the superconducting phase
by applying the probe method mentioned above. 

In $\mathcal{L}_\mathrm{CSC}$ given above, the charge $q$ is factored out by the rescaling,
$qA_{\mu}\rightarrow A_{\mu}$ and  $q{\psi}\rightarrow {\psi}$,
%\beq
%  qA_{\mu}\rightarrow A_{\mu}\,, \quad  q{\psi}\rightarrow {\psi}\,, 
%\eeq
as follows:
\beq\label{action-10}
  \mathcal{L}_\mathrm{CSC}\rightarrow {1\over q^2}\tilde{\mathcal{L}}_\mathrm{CSC}
\eeq
where $\tilde{\mathcal{L}}_\mathrm{CSC}$ is independent of $q$. This means that the probe approximation
for $\mathcal{L}_\mathrm{CSC}$ would be justified for the case of large
$q$.
Then we find that
the probe approximation for $q=2/3$ is not good. % In spite of this fact, 
However, we perform the analysis in this approximation %here under the expectation
to see what kind of results are obtained.

\vspace{.2cm}
\subsection{High temperature deconfinement phase}

First, we consider the high temperature deconfinement phase,
where the temperature is given by
\beq\label{T-AdS}
    T={5r_0\over 4\pi}\, .
\eeq 
%( see the phase diagram shown in the Fig. \ref{Phase-diagram-2-2}).
%\vspace{.3cm}
%{\bf Sol. of $\mathcal{L}_\mathrm{Gravity}$:}
The solution is known as the AdS-Schwarzschild solution, which is written as %deconfinement phase
\beq\label{Sch}
   ds^2=r^2(-f(r)dt^2+\sum_{i=1}^3 (dx^i)^2 +dw^2)+{dr^2\over r^2f(r)}\, ,
\eeq
where 
\beq
 f(r)=1-\left({r_0\over r}\right)^5\, , \quad r_0={2\over 5R_w}\, .
\eeq
Here the Sherk-Schwartz compactification is imposed in the direction $w$,
then its circle length is taken as $2\pi R_w$. % denotes the compactified length of $w$.  

\vspace{.3cm}
\noindent
{\bf Equations of probe $\mathcal{L}_\mathrm{CSC}$}

In this case the equations of motion are given as
\begin{equation}
\psi''+\left(\frac{6}{r}+\frac{f'}{f}\right)\psi'+{1\over r^2f}\left({q^2\phi^2 \over r^2 f}-{m^2}\right)\psi=0\;,
\label{eqfirst}
\end{equation}
\begin{equation}
\phi''+\frac{4}{r}\phi' -{2q^2\psi^2\over r^2 f}\phi=0\;,
\end{equation}
where we assumed $A=A_{\mu}dx^{\mu}=\phi(r)\,dt$ and $\psi=\psi(r)$.

The conformal dimension of the scalar, say $\Delta$, is related to the mass as
\beq
\Delta={1\over 2}\left(d+\sqrt{d^2+4m^2}\right)\, .
\eeq
Here we suppose that the scalar is dual to the Cooper pair, so the dimension is expected to be
$\Delta=2\times {d-1\over 2}=d-1$, which is realized for $m^2=-(d-1)$. We notice here $d=5$ and $m^2=-4$.
Then 
the asymptotic forms of $\phi$ and $\psi$ are expected as
\beq\label{Asymp1}
  \phi=\mu-{\bar{d}\over r^3}+\cdots \, , \quad \psi={J_C\over r}+{C\over r^4}+\cdots,
\quad(r\rightarrow\infty)
\eeq
where $\mu$, $\bar{d}$, $J_C$, and $C$ denote the chemical potential, charge density, source 
and the VEV of the dual operator of $\psi$, respectively.

%Here we notice that $\mu$ is supposed as the source of 

These equations are essentially equivalent to the one studied
% by Evans
in Ref.~\cite{Fadafan}.
The difference is in the dimension of the YM theory. In our case, the boundary YM theory 
is defined in 4+1 dimensional space-time with one compactified space ($w$). The effective space dimension
is however three, so there is no essential difference.
We adopt this model to analyze the confinement phase in a way parallel
to Ref.~\cite{Fadafan}, in which the vacuum solution is obtained
by the double Wick rotation of (\ref{Sch}) as shown below.

In order to avoid the singularity at the horizon ($r_0$),
the boundary conditions are given as
\beq\label{bc2}
 \phi(r_0)=0\, , \quad 
\psi'(r_0)=-{4\over 5r_0}\psi(r_0)\,,
\eeq
and the temperature is given by (\ref{T-AdS}).
%\beq\label{T-Sch}
%  T={5r_0\over 4\pi}\, .
%\eeq

\begin{figure}[htbp]%[H]
\vspace{.3cm}
\begin{center}
\includegraphics[width=7.0cm,height=7cm]{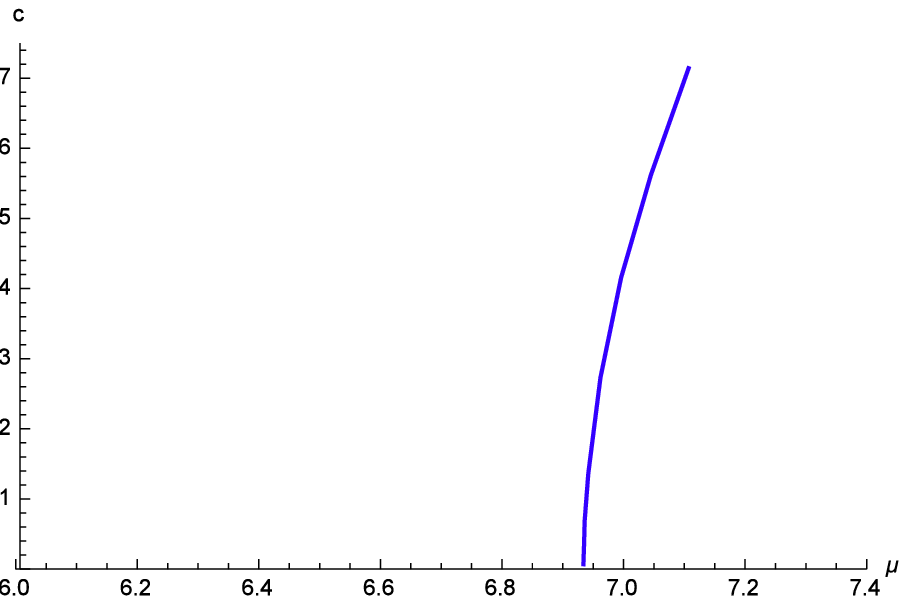}
\includegraphics[width=7.0cm,height=7cm]{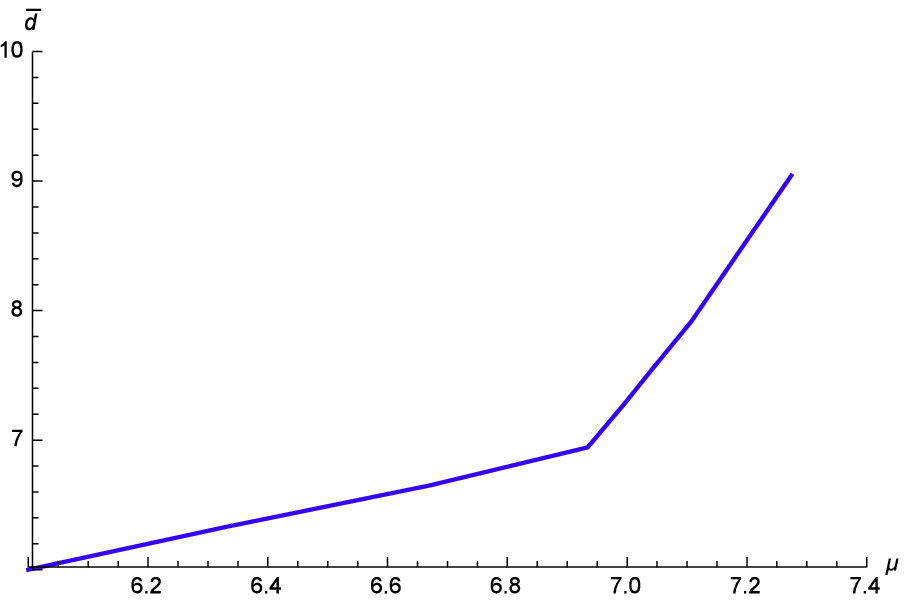}
\caption{The condensation ($C$) and the charge density ($\bar{d}$) v.s. $\mu$
in the deconfinement phase at $T=5/(4\pi)$\,.} 
\label{CS-deconfine}
\end{center}
\end{figure}

In this case, we could find the color superconducting phase
for $\mu>\mu_{\rm c}\simeq 6.9$. \footnote{The transition line
between the normal phase (b) and the color superconducting phase (d) for this case is shown
in the phase diagram in $\mu$-$T$ plane, which appears
later in Fig.~\ref{Phase-diagram-2-1}.}
This is assured by the non-trivial solution of $\psi$ for $J_C=0$
and $C\neq 0$. 
%Some of the results are shown in the Fig. \ref{CS-deconfine}.
As pointed out in Ref.~\cite{Fadafan}, there are plural solutions
(node$=0,1,2,\ldots$) for large $\mu$.
The solutions for node $\geq 1$ represent metastable vacua of the theory. 
The solution of the lowest vacuum (node$=0$) is shown in Fig.~\ref{CS-deconfine}.

As for the charge density,
we can see that it has the singularity at the transition point.
Based on Refs.~\cite{Barducci:1992jm,Kashiwa:2008ga}, such first-order
singularity (shape bend) can be understood as the singularity propagation;
it means that discontinuities appearing in particular order-parameter
can be propagated via the entropy density and/or the charge density.
Thus, the charge density also has the same order of the singularity.

%by the line between the regions (b) normal phase and super conducting (d).

\vspace{.5cm}
\subsection{Low temperature confinement phase} %;  $T\leq {5r_0\over 4\pi}$

%\vspace{.3cm}
%Sol. of $L_\mathrm{Gravity}$:

Low temperature solution of (\ref{Gravity}) is known
as the AdS soliton solution\cite{Wi,HM}, and it is obtained as
\beq\label{Soliton}
   ds^2=r^2(\eta_{\mu\nu}dx^{\mu}dx^{\nu}+f(r)dw^2)+{dr^2\over r^2f(r)}\, ,
\eeq
where 
\beq\label{Soliton-2}
 f(r)=1-\left({r_0\over r}\right)^5\, , \quad r_0={2\over 5R_w}\, ,
\eeq
and $2\pi R_w$ denotes the compactified length of $w$.  

This configuration is realized % is domminated over (\ref{Sch})
for $T\leq {5r_0\over 4\pi}$,
and the vacuum state is dual to the confinement phase. 
Namely, the line $T= {5r_0\over 4\pi}$ denotes the Hawking-Page transition line.
Under the configuration (\ref{Soliton}), we find
a linear potential between quark and anti-quark by evaluating the Wilson loop. 
%Namely the vacuum is in the confinement phase. %, and the confinement will be observed at small $\mu$.
In this case, we suppose that the condensed scalar should be a color singlet and 
the charge density %, which would be able to destroy the confinement phase,
%should be interpreted as the nucleon density since 
$\bar{d}$ is also constructed by the color singlet. 
This supposition would be correct when the chemical potential is not taken into account in the probe action
since (\ref{Soliton}) is independent of $\mu$. 
% when the density is zero or very small.

In the present case, however, the chemical potential and the charge density are contained in the probe.
Therefore, the chemical potential cannot affect the confinement background. 
%the CSC phase could appear in the confinement phase.
This implies that
it may be expected to find the phase transition to the CSC phase by solving the $\mathcal{L}_\mathrm{CSC}$ even in the case of (\ref{Soliton}). 
So it is very important to apply the model to the background (\ref{Soliton}).

\begin{figure}[htbp]%[H]
\vspace{.3cm}
\begin{center}
\includegraphics[width=7.0cm,height=7cm]{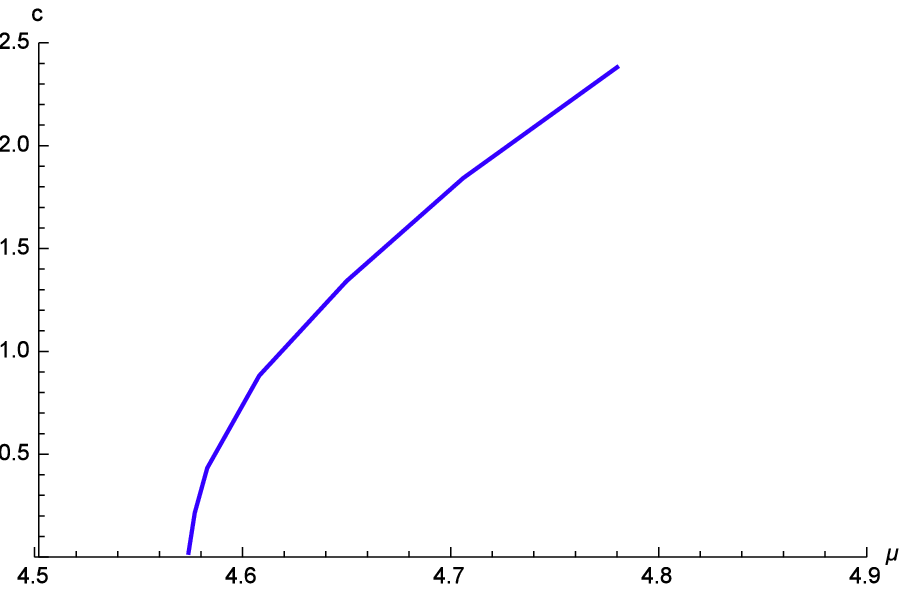}
\includegraphics[width=7.0cm,height=7cm]{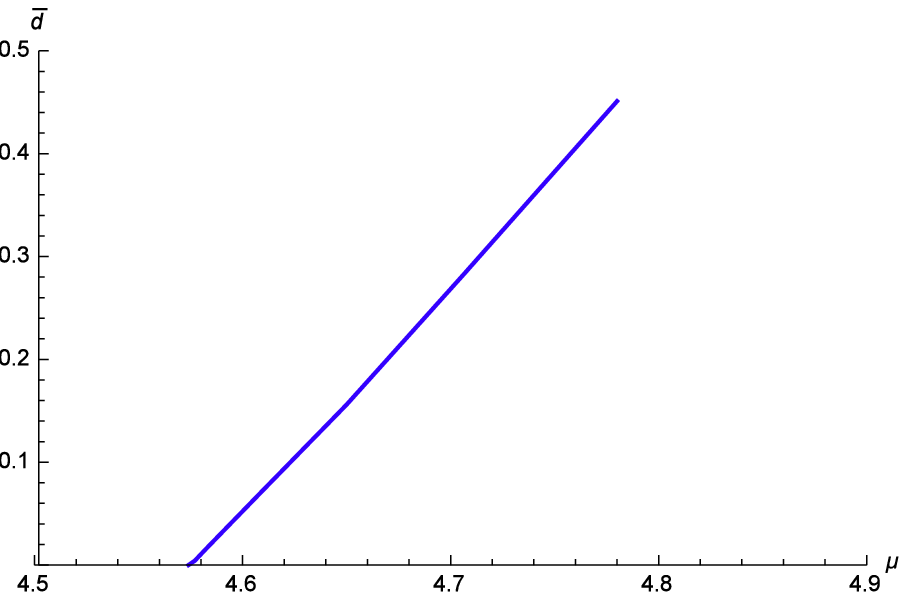}
\caption{The condensation ($C$) and the charge density ($\bar{d}$) v.s. $\mu$ in the confinement phase at 
{$T=5/(4\pi)$}\,.} 
\label{critical-mu}
\end{center}
\end{figure}

%%\begin{figure}[htbp]%[H]
%%\vspace{.3cm}
%%\begin{center}
%%%%%%\includegraphics[width=9cm,height=9cm]{Lef-Thi-2.eps}
%%\includegraphics[width=7.0cm,height=7cm]{for-AdS-Soliton-3.eps}
%%\includegraphics[width=7.0cm,height=7cm]{for-AdS-Soliton-4.eps}
%%\caption{Plots of Solution of Scalar $\psi$ and $A_0$ for $q=2/3$, $m^2=-4$.
%%$\mu=4.67793$, and $C=1.60731$ for zero node wave function $\psi$. 
%%\label{Cooper pair}}
%%\end{center}
%%\end{figure}

\vspace{.3cm}
\noindent
{\bf Equations of  probe $\mathcal{L}_\mathrm{CSC}$}

Equations of motion of $A_{\mu}$ and $\psi$ are obtained by assuming again $A=A_{\mu}dx^{\mu}=\phi(r)\,dt$ and $\psi=\psi(r)$:
\begin{equation}
\psi''+\left(\frac{6}{r}+\frac{f'}{f}\right)\psi'+{1\over r^2f}\left({q^2\phi^2 \over r^2}-{m^2}\right)\psi=0\;,
\label{eq1}
\end{equation}
\begin{equation}
\phi''+\left(\frac{4}{r}+\frac{f'}{f}\right)\phi' -{2q^2\psi^2\over r^2 f}\phi=0\;.
\label{eq2}
\end{equation}

Since $f(r)$ vanishes at $r=r_0$,
the equations (\ref{eq1}) and (\ref{eq2}) should be solved under the following conditions.
%Numerical solutions are obtained undr the boundary condition at the horizon,
\beq\label{bc1}
 \phi'(r_0)={2q^2\psi^2(r_0)\over 5r_0}\phi(r_0)\, , \quad 
\psi'(r_0)=-{1\over 5r_0}\left({q^2\phi^2(r_0)\over r_0^2}-m^2\right)
\psi(r_0)\ .
\eeq
Here, we notice the boundary condition (\ref{bc1}) allows the solution of $\phi(r_0)\neq 0$.

\vspace{.3cm}
As expected, we could find non-trivial solutions of $\psi$ with $J_C=0$ and $C\neq 0$ for $\mu>\mu_{\rm c}^\mathrm{conf}\simeq 4.7$ with $r_0=1$. 
%An example of 
For such  non-trivial solutions, % for $\psi$ and $A_0=\phi$ is shown in the Fig. \ref{Cooper pair}.
%Also we show The 
the $\mu$ dependence of $\bar{d}$ and $C$ in the confined phase are shown in Fig.~\ref{critical-mu}.

\begin{figure}[htbp]%[H]
\vspace{.3cm}
\begin{center}
\includegraphics[width=7.0cm,height=7cm]{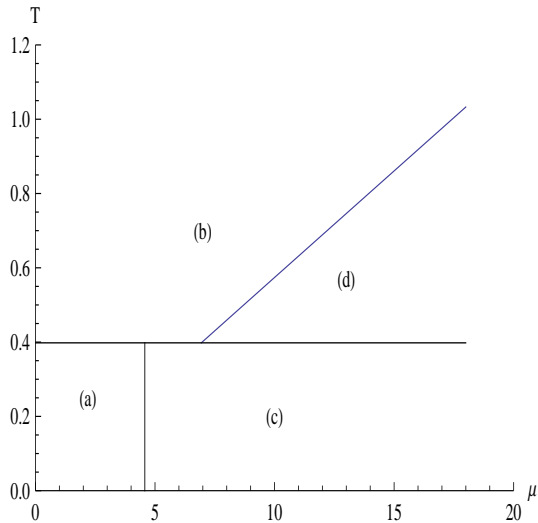}
\caption{ {Phase diagram given by the probe approximation for $q=2/3$, $m^2=-4$. 
The regions (b) and (d) represent the normal and CSC phase
for AdS-Schwarzschild de-confinement background. The critical line between (b) and (d) is given by
$T=0.058\mu$. For AdS soliton confinement background, (a) and (c) represent the 
normal and CSC phase with the critical line $\mu=4.7$ which is independent of $T$.}
 %Here, $q=2/3$, $m^2=-4$. % or $\Delta=4=2\times 2$.
%Right: Phase diagram when (d) and (c) in the left are identified. This has the topologically same structure with the one of Fig.~\ref{Phase-diagram-2}. 
\label{Phase-diagram-2-1}}
\end{center}
\end{figure}

From the present analysis, we can draw the phase diagram
in the $\mu$-$T$ plane,
as shown by Fig.~\ref{Phase-diagram-2-1}.
We notice the existence of the critical line
between the regions (a) and (c).
{As a result, there appear two superconducting phases (c) and (d).
It is an interesting point how they are different from each other.}

\vspace{.3cm}
{In the phase (c), the bulk represents the confinement phase. This phase would not be 
compatible with %Then 
the existence of the 
VEV of the diquark operator and of the finite charge density.} % would not be compatible.}
The reason why the CSC phase appears in the confinement bulk background
%, which is expressed by the AdS soliton solution,
%as found in the deconfinement backgrou
{may}
be reduced to the probe approximation used in this section. {As explained above through 
the equation (\ref{action-10}), which is given by
the rescaling $q A_{\mu}\to A_{\mu}$ and $q\psi\to \psi$,
%(\ref{action}) is rewritten as
%\beq
%\mathcal{L}=\mathcal{L}_\mathrm{Gravity}+{1\over q^2}\mathcal{L}'_\mathrm{CSC}\, , \label{action-2}
%\eeq}
%where $\mathcal{L}'_\mathrm{CSC}$ does not include $q$.
%This implies that 
the probe approximation is useful for large $q$. However, 
$q(=2/N_c)$ should be very small in the present case since the holographic approach is available for large $N_c$. }

{An alternative idea to support the probe approximation is to suppress %reduce the origin of 
$\mathcal{\tilde{L}}_\mathrm{CSC}$ by a small number
of the flavor branes, $N_f$, which would appears as the prefactor as %of above equation can be written as
\beq
\mathcal{L}=\mathcal{L}_\mathrm{Gravity}+{N_f\over q^2}\mathcal{\tilde{L}}_\mathrm{CSC}\, . \label{action-3}
\eeq
Then we can say the validity of the probe approximation for $N_f/q^2 <<1$. This is however impossible since $N_f>1$. 
Therefore, it seems difficult to support the validity of the probe approximation}
%to accept the result of this section.}
{Then, we should improve the probe approximation by considering the back-reaction from the terms of 
$\mathcal{L}_\mathrm{CSC}$ as performed in the next section.}

\section{Superconductor in the back-reacted background}
\label{sec:backreaction}

In the previous section, we find a phase transition to a color superconducting phase even in the confinement background of the 
theory. 
Why has such a transition been observed in the confinement vacuum? 

As mentioned above,
this is because of the application of the probe approximation without any back-reaction from the matter field part of the model.
%existence of the chemical potential and the charge density in the probe action. 
The superconducting phase 
is observed in the large $\mu$ region, $\mu \geq\mu_{\rm c}$. 
We expect that, in this region, the confinement force is suppressed 
{by the effect of the chemical potential and the charge density} so that
%a strong electric force and
the deconfinement phase might be realized.
%In other words, this implies the occurrence of the transition to the deconfinement phase simultaneously. 
%For small $\mu$, the confinement is preserved, and the superconducting behavior cannot be seen.}
In order to make the situation clear, we should take account of the back-reaction of the probe action to the
gravity part. 
This is performed here according to the way given in Ref.~\cite{Basu}.

At first, the basic background configuration is set
by taking account of the $U(1)$ gauge part into the gravity part.
In other words, the back-reaction from the $U(1)$ gauge part
to the bulk gravity is fully considered since
it becomes important in the region, $\mu \geq \mu_{\rm c}$.

\vspace{.3cm}
Now, we set the bulk background configuration
by solving the following action\cite{Iqbal},
\beq\label{B}
S_\mathrm{Bulk} = \int d^{6} x \sqrt{-g} \left\{ {\cal R} + {20 \over L^2}- {1 \over 4} F^2 \right\}\, .
\eeq
The action leads to the following three solutions. % as the model (A). 

\begin{itemize}
\item[(1)]AdS soliton (confinement phase)

The solution is given by the background metric (\ref{Soliton}),  (\ref{Soliton-2}), 
and the constant potential,
\beq\label{constant}
  A_0=\phi=\mu \, .
\eeq
%For this solution, the probe should be taken as $\mathcal{L}_{SCS}$
%  as in the case (A) since $U(1)$ gauge part is neglected

\item[(2)]AdS-Schwarzschild (deconfinement phase)

The solution with the background (\ref{Sch}) is compatible
with the same type of a constant potential as (\ref{constant}).

\item[(3)]Reissner-Nordstrom (RN) (deconfinement phase) 

By considering the $U(1)$ potential $A_0$ with a finite charge density,
a charged black hole solution (see \textit{e.g.\/}, Refs.~\cite{Iqbal}, \cite{Iqbal:2010eh})
is obtained\footnote{
The notation is taken following Ref.~\cite{Basu}.
}:
\beq\label{RN}
   ds^2=r^2\left(-g(r)dt^2+\sum_{i=1}^3 (dx^i)^2 +dw^2\right)+{dr^2\over r^2g(r)}\, ,
\eeq
\beq
 g(r)=1-\left(1+{3\mu^2\over 8 r_+^2}\right)\left({r_+\over r}\right)^5+{3\mu^2r_+^6\over 8r^8}
\label{RN2}\, ,
\eeq
\beq\label{RN3}
  A_0=\phi=\mu\left(1-{r_+^3\over r^3}\right),
\eeq
where $r_+$ denotes the horizon of the charged black hole,
and the Hawking temperature is given by
\beq\label{T-RN}
  T={1\over 4\pi}\left(5r_+-{9\mu^2\over 8r_+}\right)\, .
\eeq
\end{itemize}

{Here we notice on the setting of the parameter $q$. As mentioned above, 
by rescaling as $q A_{\mu}\to A_{\mu}$ and $q\psi\to \psi$,
(\ref{action}) is rewritten by Eq.(\ref{action-10}). Then the equations of motion derived from $\mathcal{\tilde{L}}_\mathrm{CSC}$ for 
the rescaled
$\phi$ and $\psi$ are independen of $q$. However, $q$ appears in the equations of motion when 
we take into account of the backreaction
to the background determined by the
garivity since $q$ remains as the prefactor of $\mathcal{\tilde{L}}_\mathrm{CSC}$. Due to this fact, 
in this section, we solve the equations
without any rescaling mentioned above.}

\vspace{.3cm}
\subsection{Phase diagram before adding scalar}

Before adding the scalar field, we compare the free energies of three types
of vacua which arise from the background configurations, those are,
%For these vacuum, we show the phase diagram 
AdS black hole (BH), AdS soliton (Soliton) and  Reissner-Nordstrom black
hole (RN) given by (\ref{Sch}), (\ref{Soliton}) and (\ref{RN}),
respectively.

\begin{figure}[htbp]%[H]
\vspace{.3cm}
\begin{center}
\includegraphics[width=7.0cm,height=7cm]{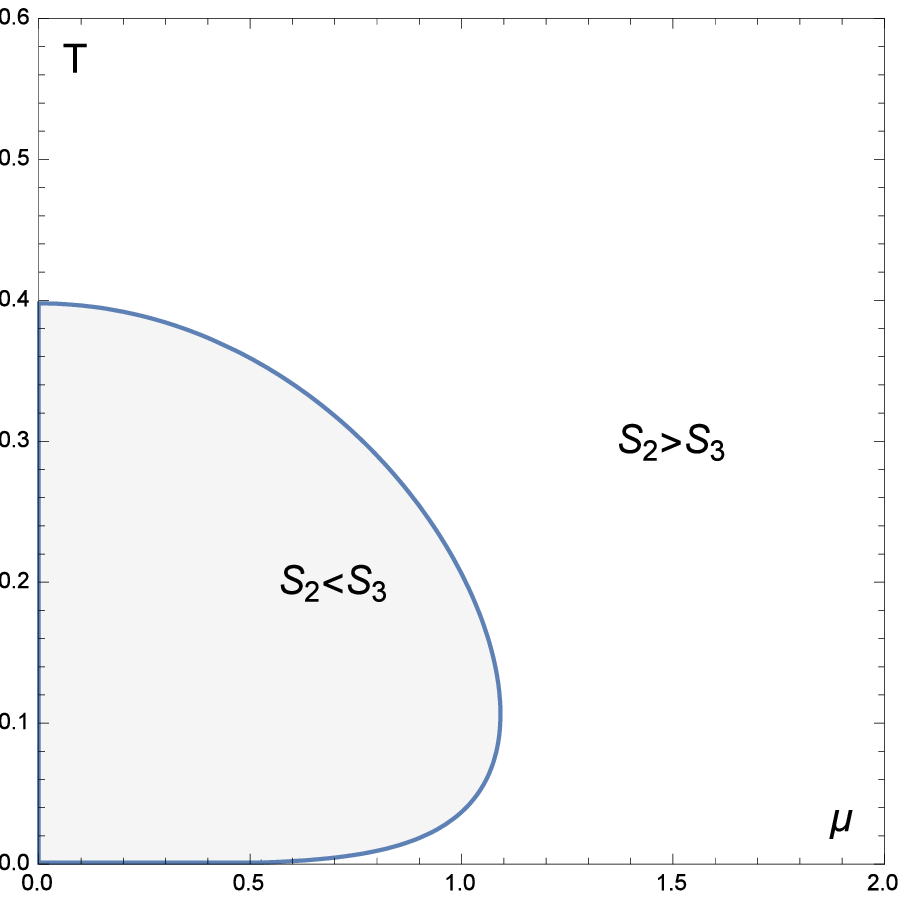}
\includegraphics[width=7.0cm,height=7cm]{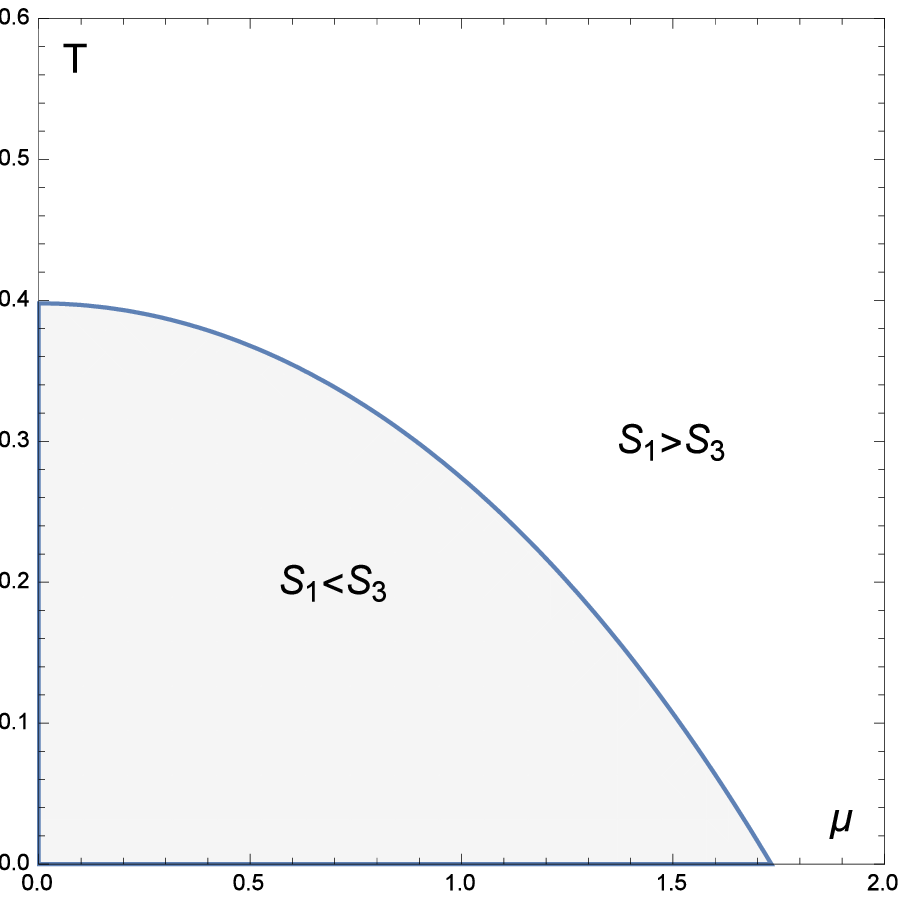}
\caption{Comparison among the actions ($S_2$ v.s. $S_3$ and $S_1$ v.s. $S_3$)
\label{Phase-diagram}}
\end{center}
\end{figure}

%\vspace{.3cm}
%\subsection

%First, we show the phase diagram in the $\mu -T$ plane by comparing the actions of three possible background
%solutions mentioned above, (1), (2), and (3). 
Action densities for the three solutions are given as
\bea
S_1/V_3 &\equiv S_\mathrm{(Soliton)}/V_3 &= -r_0^5{4\pi\over 5r_0}{1\over T}\, \\
S_2/V_3 &\equiv S_\mathrm{(BH)}/V_3 &= -r_0^5\left({4\pi\over 5r_0}\right)^2\, \\
S_3/V_3 &\equiv S_\mathrm{(RN)}/V_3 &= -r_+^5\left(1+{3\mu^2\over 8r_+^2}\right){4\pi\over 5r_0}{1\over T}\, 
\eea
where $V_3=\int\!dxdydz$. We notice that the temperature $T$ for RN and BH are given by (\ref{T-RN}) and (\ref{T-AdS}), respectively.
In the $\mu$-$T$ plane, we find the phase diagram by choosing the lowest action density among the above three.
% (1) Soliton, (2) BH, and (3) RN.
We find that the solution of BH is not realized when the solution of RN is added to compare.

The difference of the actions are written as,
\beq
  (S_i-S_j)/V_3=N_{ij}(X_{ij}^5-Y_{ij}^5)
\eeq
where $i$ and $j$ run form 1 to 3, and
\beq
 N_{23}=N_{13}={4\pi\over 5r_0T}\, , \quad X_{13}=X_{23}=r_+\left(1+{3\mu^2\over 8r_+^2}\right)^{1/5}\, ,
\eeq
\beq
   Y_{23}=r_0\left({4\pi T\over 5r_0}\right)^{1/5}\, , \quad Y_{13}=r_0=1\, .
\eeq
Then, which action is larger is found by the plot of $X_{ij}-Y_{ij}$, which are shown in Fig.\ref{Phase-diagram}.

%\beq
% \Delta S_{23}=r_p\left(1+{3\mu^2\over 8r_+^2}\right)^{1/5}-{}
%\eeq

In the left hand side of Fig.~\ref{Phase-diagram},
there is a region where $S_{2}<S_{3}$.
However, we find easily $S_{2}>S_{1}$ in this region.
In fact, $S_{1}<S_{2}$ for $T<5r_0/4\pi$ at any $\mu$.
Then the phase diagram is constructed by the phase represented
by Soliton and RN backgrounds,
and it is given by the right hand side of Fig.~\ref{Phase-diagram},
which is separated by the phase of Soliton and RN.
The critical curve in Fig.~\ref{Phase-diagram-2}, which is shown in the next subsection \ref{subsection-3-2}, is given by
\beq
  X_{13}=1\,.
\eeq

\vspace{.3cm}
We notice the following point.
In the previous section, 
the gauge field part is treated as the probe.
It is not used for constructing the background metric.
As a result, the phase diagram is given by comparing $S_{2}$ and $S_{1}$.
In this case, the parameter space is separated to two phases,
corresponding to the Soliton solution and the BH one,
and the critical line is given by $T=5r_0/4\pi$. 
Therefor the phase diagram in the previous section is largely changed
after adding the $U(1)$ gauge part in the equations of motion to be solved. %probe which leads to the color superconductor.
%the analysis with respect to the color conductor,
%which has been given in the above subsection.

%\newpage
\subsection{Phase diagram after adding scalar}
\label{subsection-3-2}

%In the present case,
The phase diagrams shown in Fig. \ref{Phase-diagram}
would be changed due to the appearance of the color superconducting phase,
which could be found by the scalar field condensation.
%as a probe in the two bulk background configurations, Soliton and RN. 
We should remember that the backgrounds considered
here are obtained by taking into account of the back-reaction of $U(1)$ gauge part. 
The back-reaction of the scalar is also taken into account when we solve its equation of motion.
{In this case, we must solve the equations of back-reacted metric and gauge field, 
and the equation of the scalar
at the same time.}
Although it is a straightforward but hard work to solve those simultaneous equations, % proceed this calculation, 
we can find a phase diagram after solving them.
%with a critical curve
%separating the color normal and superconducting phases when it exists.

{On the other hand, it would be possible to find the critical curve without solving the full equations. 
%it is enough to obtain a new phase diagram. 
This economical method has been proposed in Ref.~\cite{Basu}.}
%as explained below,
%and then the critical curve is studied here according to this method.} 
%\kg{The idea is explained below.}
When a superconducting phase exists, % in the RN background,
there is a solution for the scalar %with the asymptotic form (\ref{Asymp1})
with $J_C=0$ and a finite $C$ for $\mu > \mu_{\rm c}^\mathrm{(B)}$ at a finite $T$. And, for
$\mu \to \mu_{\rm c}^\mathrm{(B)}$, $C$ approaches zero. \footnote{Here we suppose the order parameter
$C$ has no gap at the transition point as seen in the previous sectin for the probe approximation. }
%since the strength of the scalar is very small near the critical line.
Then the back-reaction from the scalar to the bulk configuration becomes negligible near the critical point.
{This means that the back-reactions to both the metric and gauge field disappear.}
This implies that the critical line can be obtained by solving the equation of motion of only the scalar which is
treated as a probe in the two vacuum configurations given above. Then the task to find the critical line
is to solve the equation of the scalar field in the given background of $S_\mathrm{Bulk}$ denoted in (\ref{B}). 

\vspace{.3cm}
\noindent{\bf Deconfinement phase}

At first we consider the equation of the motion in the RN background, (\ref{RN}) -- (\ref{RN3}). It is given as
\beq\label{critical-s}
  \psi''+\left(\frac{6}{r}+\frac{g'}{g}\right)\psi'+{1\over r^2g}\left({q^2\phi^2 \over r^2 g}-{m^2}\right)\psi=0\;.
\eeq
%in the background (\ref{RN}) -- (\ref{RN3}). 
The boundary condition to be imposed is
\beq
  \psi'(r_+)={m^2\over 5r_+-{9\mu^2\over 8r_+}}\,\psi(r_+)\ .
\eeq

The equation (\ref{critical-s}) includes parameters, $q$, $\mu$ and $r_+$. Here the temperature is
expressed by $r_+$ and $\mu$
as (\ref{T-RN}). Since $T$ should be positive, then we find the following constraint for $\mu$,
\beq\label{Bound-mu}
  0\leq {\mu\over r_+}\leq {\sqrt{40}\over 3}\, .
\eeq
Under this constraint, we searched for a solution with $J_C=0$ and $C\neq 0$. 
{And, within the error of our numerical calculation, we get to the conclusion that there is no such a solution for $q=2/N_c\leq 1$. }
Since our concern is in the case
of $N_c=3$ (or $N_c=2$), we cannot say that there exists %we have found 
a color superconducting phase in 4D Yang-Mills
theories with quarks. 

\vspace{.3cm}
{According to a view} pointed out in Ref.~\cite{Fadafan}, 
the reason why the solution leading to the condensation of the Cooper pair is not found in the present case
is explained by considering the effective mass of the scalar,
$m_{\rm eff}$. 
The input mass $m^2=-4$ satisfies
the Breitenlohner Freedman (BF) bound ($-25/4<m^2$) in the 5+1 dimensional AdS space-time.
However, as shown below, %seen in the previous section, 
the mass is effectively
suppressed by the coupling to the gauge potential $\phi$. 
Noticing that Eq. (\ref{critical-s}) is rewritten as $(\Box_{\rm RN}-m^2_\mathrm{eff})\psi =0$,
where $\Box_{\rm RN}$ denotes the Laplacian in the RN background,
$m_{\rm eff}$ can be read 
%Here we notice the effective scalar mass, which is defined 
%from (\ref{critical-s}) 
as follows: 
\beq
  m^2_\mathrm{eff}=m^2-\Delta m^2\,, \quad   \Delta m^2\equiv {q^2\phi^2\over r^2 g}\,.
\eeq 
Then we consider that the necessary condition to destabilize the scalar field
and to condense in the vacuum is to break the BF bound
for $m_{\rm eff}$. It is given by $ m^2_\mathrm{eff}< -{25\over 4}$ or equivalently by
%and it is given by
\beq\label{Condi-1}
   \Delta m^2> {9\over 4}\, .
\eeq
Here we notice that $\Delta m^2$ is $r$-dependent and it is rewritten by using (\ref{RN2}) and (\ref{RN3}) as 
\beq\label{BF-1}
   \Delta m^2= {q^2}F(x,\tilde{\mu})\, , \quad \tilde{\mu}={\mu\over r_+},
\eeq
\beq
  F(x,\tilde{\mu})={x^2(1-x^3)^2\tilde{\mu}^2 \over 1-(1+{3\tilde{\mu}^2\over 8})x^5+{3\tilde{\mu}^2\over 8} x^8 },
\eeq
where $x=r_+/r$.

\vspace{.3cm}
%******************************
\begin{figure}[htbp]%[H]
\vspace{.3cm}
\begin{center}
\includegraphics[width=9.0cm,height=7cm]{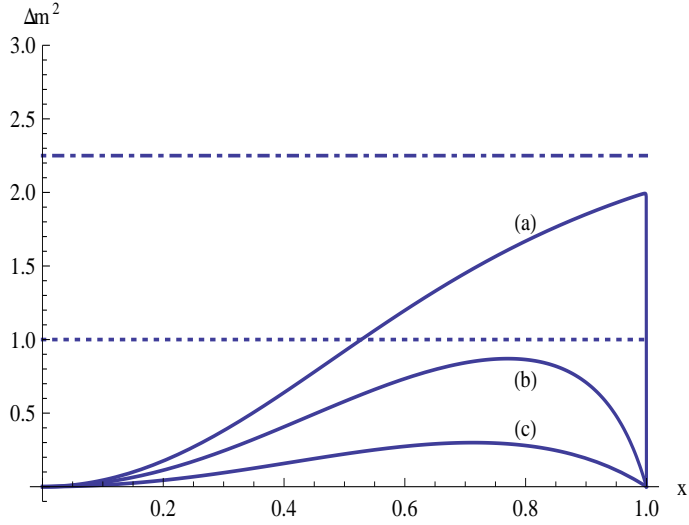}
\caption{%Left: Phase diagram for AdS-Schwarzschild and Right: 
The value of $\Delta m^2$ for $q=1$, (a) $\tilde{\mu} ={\sqrt{40}\over 3}$ ,  (b) $\tilde{\mu} =0.8\times{\sqrt{40}\over 3}$,
and {(c)} $\tilde{\mu} =0.5\times{\sqrt{40}\over 3}$.
The horizontal dot-dashed and dotted lines show 9/4 and 1.0, the bound to break the BF bound for $\Delta m^2$
in the AdS$_6$ and infrared AdS$_2$.
}
 %\label{Phase-diagram-2}
 \label{RN-BF}
\end{center}
\end{figure}

The factor $F(x,\tilde{\mu})$ increases with $\tilde{\mu}$ at any $x$. The value of $\tilde{\mu}$ 
is bounded from above as shown by (\ref{Bound-mu}). So we can find the maximum form of $F(x,\tilde{\mu})$
as $F(x,\sqrt{40}/3)$ as shown by the curve (a) in Fig.~\ref{RN-BF}. Further the maximum value of this
function is found as $F(x,\sqrt{40}/3)<1.993 \simeq 2$. Then we are led to
%
%In Fig.~\ref{RN-BF},  $\Delta m^2$ is plotted for various values of $\tilde{\mu}$ below its upper bound, which is
%given by (\ref{Bound-mu}). %, \textcolor{red}{where $q=1$.} 
%From $F(x,\tilde{\mu})<1.993 \simeq 2$, we find
\beq\label{BF-2}
  0<\Delta m^2 <2 q^2={8\over N_c^2}
\eeq
for all region of $r (\geq r_+)$.
%In the right hand side equation, the following charge assignment is used, 
The above equality comes from $q={2\over N_c}$
%Here the charge of the scalar field is taken as 
%\beq
% q={2\over N_c}\, ,
%\eeq
since we assume that the scalar is dual to the diquark operator. 

From (\ref{BF-2}) and (\ref{Condi-1}),
we have the condition, that $\Delta m^2$ breaks the BF condition at some point of $r$,
as %follows %we have
\begin{align}
\frac{9}{4} < \Delta m^2 < \frac{8}{N_c^2}\,,
\end{align}
and thus we obtain
\beq
      N_c< \frac{4\sqrt{2}}{3} \simeq 1.89 \, .
\eeq
%to break the BF condition at some point of $r$. %locally in the $r$ direction. 
{This implies that} 
%Then, as mentioned above, 
it is impossible to see the scalar
condensate with reasonable values of $N_c$,  namely in the region $N_c\geq 2$, in the present holographic model.
As a result, {we can say that} there is no CSC phase in the RN background.
%when the parameters are set as the reasonable values.

\vspace{.3cm}

In arriving at the above result, we must notice the following
%important 
comments.

%\vspace{.3cm}
\noindent {\bf (C1)} While %we give 
a necessary condition for the condensation of the scalar is given above,
we did not say
about the sufficient condition. It is shown by a simple example studied in the previous section.
For the case of the confinement phase, from Eq. (\ref{eq1}) we can set as
\beq\label{confine-mass}
    \Delta m^2= {q^2\phi^2 \over r^2}=q^2{\mu^2\over r_0^2}\left({r_0\over r}\right)^2\, . %, \quad  x={r_0\over r}\, .
\eeq
For $r_0=1$, the necessary condition of BF bound breaking is given by $q\mu>1.5$. However we need
$q\mu>3.01$ \footnote{The value of $q\mu$ is obtained by our numerical estimation.}
to find the CSC phase. %critical point
%of getting the superconductivity is $q\mu>3.01$. 
This implies that an enough 
%So a 
wide region of $r$, where BF bound is broken, should be needed
as a sufficient condition of the scalar condensation. This statement for the sufficient condition
is available in other cases.

\begin{figure}[htbp]%[H]
\vspace{.3cm}
\begin{center}
\includegraphics[width=9.0cm,height=7cm]{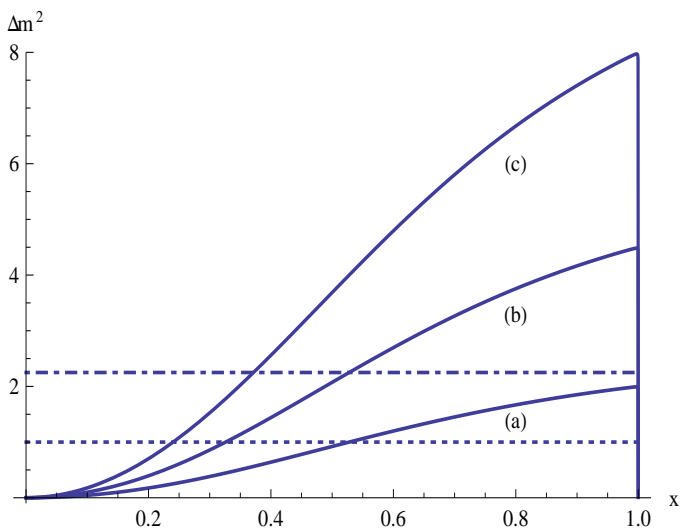}
\caption{%Left: Phase diagram for AdS-Schwarzschild and Right: 
The value of $\Delta m^2$ for $\tilde{\mu} ={\sqrt{40}\over 3}$ and(a) $q=1$, (b) $q=1.5$,
and {(c)} $q=2$.
The horizontal dot-dashed and dotted lines show 9/4 and 1.0, the bound to break the BF bound for $\Delta m^2$
in the AdS$_6$ and infrared AdS$_2$. For $q\geq 1.5$, the CSC phase has been found.
}
 %\label{Phase-diagram-2}
 \label{RN-BF2}
\end{center}
\end{figure}

In Fig. \ref{RN-BF2}, some examples of $m_{\rm eff}$ in the deconfinement phase are shown at $T=0$.
In this case, the sufficient condition is given as $q>1.5$, and its $m_{\rm eff}$ is shown by the curve (b) of Fig. \ref{RN-BF2}.

\begin{figure}[htbp]%[H]
\vspace{.3cm}
\begin{center}
\includegraphics[width=7.0cm,height=7cm]{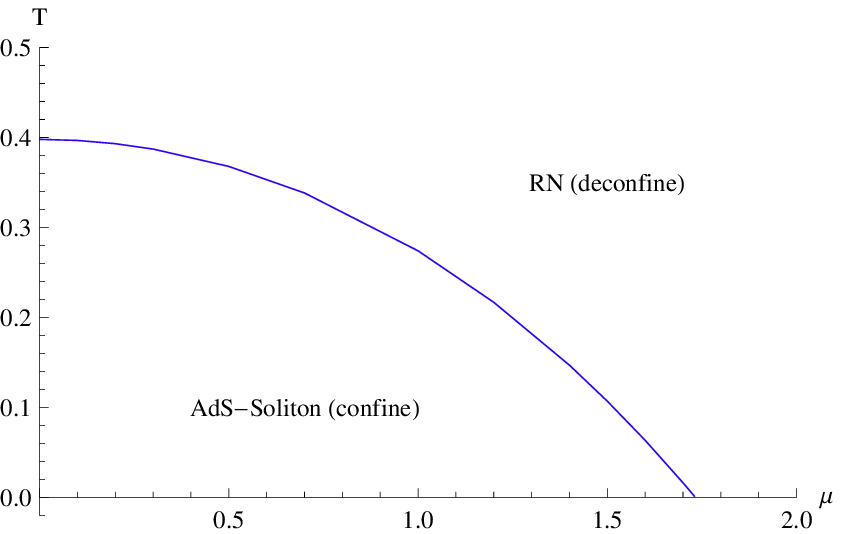}
\includegraphics[width=7.0cm,height=7cm]{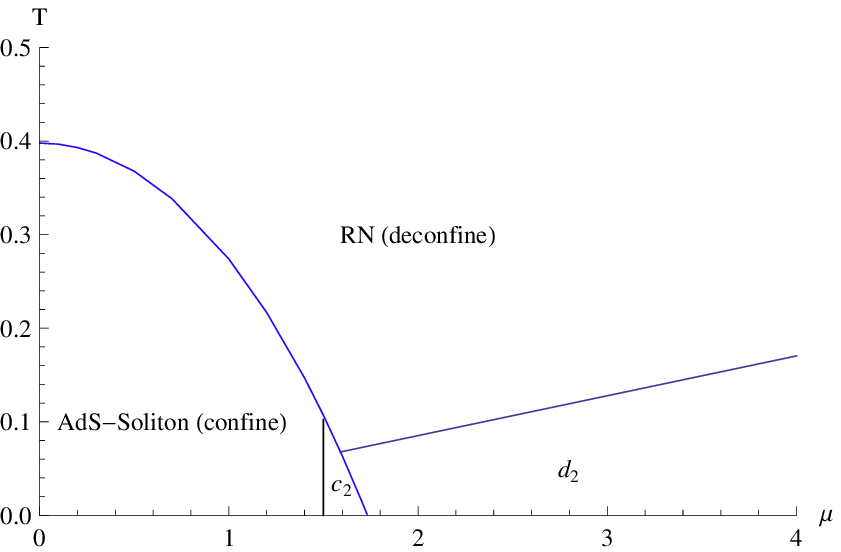}
\caption{Left: The phase diagram for the case of the back-reacted background for $N_c\geq 2$. 
There is no CSC phase. The critical curve between the deconfinement and 
the confinement phases, which are denoted by RN(deconfine) and AdS-Soliton(confine) respectively,
is given in the right figure of  Fig.~\ref{Phase-diagram}.
Right: The phase diagram for $N_c=1$. The regions ($c_2$) and ($d_2$) denotes
the CSC phases, and the critical line in the RN represents $T/\mu=0.0426$.
The vertical line in confinement phase denotes $\mu=1.5$.}
\label{Phase-diagram-2}
\end{center}
\end{figure}

\vspace{.3cm}
{\noindent {\bf (C2)}
In the case of the deconfinement phase, the curve (a) in Fig.\ref{RN-BF} is lower than the bound for all $x$.
This curve is given for $q=1$ ($N_c=2$). On the other hand, 
we can find a CSC phase with the critical line $T/\mu=0.0426$ for $q=2$ ($N_c=1$) (see 
the curve (c) of Fig. \ref{RN-BF2} and
the right hand figure of Fig. \ref{Phase-diagram-2}). 
Then, in this case, the magnitude of
the scalar 
charge is sufficient to genarate the condensation. However, in this case, we have $N_c=1$, which 
means that the dual theory is $U(1)_c$ gauge theory. So it is unrealistic
in the present bottom up model as the SYM theory and also from the holographic set up.}

\vspace{.3cm}
%******************************

%*********************************

%Left of Fig.\ref{Phase-diagram-2}; (a) AdS-soliton, Confinement phase, (b) AdS-Schwarzschild, deconfinement phase, (c) deconfinement and color super-condoctor
%phase.

\vspace{.3cm}
\noindent {\bf (C3)}
We notice another scalar mass bound which has been examined in \cite{Basu}.
Near the horizon of the RN black hole solution, we find a geometry AdS$_2\times R^4$ \cite{Hartnoll:2008kx,Iqbal:2010eh} where
the radius of AdS$_2$ is given by $L/\sqrt{20}$. 
Then the new BF bound near this AdS$_2$ geometry is given
by $m_\mathrm{eff}^2>-5$. This bound is broken for $\Delta m^2 >1$.
From Fig.~\ref{RN-BF}, the curve (a) is over the bound value in a long range of $x$. This implies
a possibility of the solution of the scalar which indicates the Cooper pair condensation at very small temperature
as found in \cite{Basu} for a neutral scalar with $m^2 < -5$. 
However we cannot find such a solution in the present model. So 
one may consider that the curve (a) is not 
yet sufficient to generate CSC phase as mentioned in the above comment (C1). Alternate reason of not finding
the CSC phase may that we are not considering the neutral scalar but a charged one. Since the operators dual to the scalars
are different from each other, then our result might be compatible with the one of Ref.~\cite{Basu}. 
In any case, this point is an open problem.

\vspace{.3cm}
\noindent{\bf Confinement phase}

In the confinement region, the equations to be solved are given as
\begin{equation}
\psi''+\left(\frac{6}{r}+\frac{f'}{f}\right)\psi'+{1\over r^2f}\left({q^2\phi^2 \over r^2}-{m^2}\right)\psi=0\;,
\label{eq1-RN}
\end{equation}
\begin{equation}
f=1-\left({r_0\over r}\right)^5\, , \quad \phi=\mu\, ,
%\phi''+\left(\frac{4}{r}+\frac{f'}{f}\right)\phi' -{2q^2\psi^2\over r^2 f}\phi=0\;,
\label{eq2-RN}
\end{equation}
with the following condition at $r=r_0$, %since $f(r_0)$ vanishes there,
%Numerical solutions are obtained under the boundary condition at the horizon,
\beq\label{bc1-RN-2}
% \phi'(r_0)={2q^2\psi^2(r_0)\over 5r_0}\phi(r_0)\, , \quad 
\psi'(r_0)=-{1\over 5r_0}\left({q^2\phi^2(r_0)\over r_0^2}-m^2\right)
\psi(r_0)
\eeq

In this case, the effective scalar mass has the same form with the case considered in the previous section.
Hence  $\Delta m^2$ is given by (\ref{confine-mass}). Thus the instability would be seen locally for
$q\mu >1.5$ when we set as $r_0=1$. However the sufficient condition to find CSC phase is $q\mu >3.01$
as mentioned above. This implies $\mu >3.01$ for $q=1$. This region of $\mu$ is out of the confinement phase, 
$\mu<1.73$. So we can not find CSC phase also in the confinement background.
Noticing $q=2/N_c$, the CSC phase cannot be found for $N_c\geq 2$.

\vspace{.3cm}
We should notice that the CSC phase is realized for $q=2$ ($N_c=1$) since the sufficient condition is given
as $\mu >3.01/q=1.5$ which is within the confinement region. The phase diagram for $q=2$ case is given
in Fig. \ref{Phase-diagram-2}.

\vspace{.3cm}
%\noindent{\bf Conclusion}

This section can be summarized as follows:

(1) For $N_c\geq 2$,  we can say that there is no CSC
phase in the deconfinement and also in the confinement phases. Then, we need not to change the phase diagram 
even if an appropriately charged scalar is added by taking into account of the back-reaction to realize the color superconductivity
in the SYM theory. % \textcolor{red}{for $N_c\geq 2$.}  

(2) For $N_c=1$, a phase diagram with the CSC phase is obtained, but the setting of $N_c=1$ is unrealistic
since it is incompatible with the holography. So it is hard to accept the results obtained in this case.

\vspace{.3cm}
{We must remember that the CSC phase searched in this section is the one realized by the phase transition
without the gap of the order parameter $C$. In this paper, we concentrated on this type of phase transition since the same type 
transition
is found in the case of the probe approximation as shown in the previous section 2.}

%---------------------------------------------------------------------------

%--------------------------------------------------------------------------
\section{Summary and Discussion}

We have studied a possibility of the CSC phase in the SYM theory by using a bottom-up holographic model
which is constructed by the gravity and a simple matter action composed of a $U(1)$ gauge field and a charged scalar.
%\kg{with the mass of $m^2=-4$}.
The time component of the gauge field gives a finite chemical potential $\mu$ of the baryon number and its 
density in the vacuum of the SYM theory. %The dual of the vacuum state is given as the solution of the bulk gravity. 
The mass and the charge of the scalar give the conformal dimension and the baryon number 
of the composite operator of the dual SYM theory. It is chosen as a scalar which is dual to
the diquark operator. According to this holographic setting,
the equations of motion of the system are solved, and %we searched for
the CSC phase of the SYM theory is searched. 

Using this model, %In the setting mentioned above
at first,
a probe analysis has been applied to the two vacuum states, the confining and deconfinement phases.
The probe action is composed of the $U(1)$ gauge field and the charged scalar.
The bulk configurations of these two phases
%are given by %as the solutions of 6D gravity, namely 
%the AdS soliton and the AdS black hole solution, respectively. They 
are therefore independent of $\mu$ since the chemical
potential and the charge density belong to the probe. In this case,
we find the CSC phase in each vacuum when the value of $\mu$ exceeds a
critical point observed in each phase. %, $\mu_{\rm c}^{\rm conf}$ or $\mu_{\rm c}^{\rm deconf}$.
In any case, this transition causes the breaking of the gauge 
symmetry since the VEV of the diquark operator is not a color singlet except for the case of $N_c=2$. 
In this sense, it might be
interesting to study %the relation between 
the color superconductors in the 
holographic Higgs branch \cite{FMPT,Apreda:2005yz,Guralnik:2004ve,Erdmenger:2005bj} %Rozali:2012ry,Guralnik:2004ve,Guralnik:2005jg,Guralnik:2004wq,Erdmenger:2005bj}.
The approach in this direction would be given elsewhere.

%Furthermore,
In any case, 
this CSC transition implies the transition from the confinement to the deconfinement phase at the same time.
%Therefore, the property of the vacuum which was given by the bulk gravity is changed by the probe.
%, namely from confinement tothe deconfinement. 
The reason why this curious phenomenon occurs would be that 
the probe approximation is used in the situation
where this approximation may not be useful.
In fact, here the holographic approach 
%the charge 
is set for $q=2/N_c$ with large $N_c$, so $q$ is small. On the other hand, large $q$ is needed
for the validity of the probe approximation.

In order to {improve the results the probe approximation, we consider the back-reactions. At first, we}
reset the vacuum. %such that it 
It is given as a solution of the gravity with the full back-reaction
from the $U(1)$ gauge field. % part of the action.
As a result, %the phase diagram determined by using the back-reacted 
the solutions and the phase diagram, %of the bulk gravity
which is given by the two phases, confinement (AdS soliton) and deconfinement (Reissner-Nordstrom black hole),
have been largely changed and are depending on $\mu$. % by the effect of the chemical potential coming from the probe. 
By adding the scalar in these vacuum configurations, the superconducting phase 
%In this case, the remained scalar part
has been searched. {Here the back-reaction of the scalar is also considered,}
and we arrive at the conclusion that
%could not find the 
there is no CSC phase in both the confinement and deconfinement
phases with the reasonable parameters of the theory.

We should however notice that we find a region of $r$ where
the BF bound is broken, for $N_c=2$, in the confinement phase.
But enough instability to generate the CSC phase
cannot be obtained in this case.\footnote{
As a related example, another kind of phase transition
in the confinement phase has been studied by using a different form
of probe composed of $\mathrm{D}8\bar{\mathrm{D}8}$ branes,
in which the baryon chemical potential is considered \cite{GKTTT}}
There are some studies of the CSC phase via two color lattice QCD
(QC$_2$D)
%%%%%%%%%%%
since the fermion determinant is pseudo-real and then the sign problem 
vanishes. %anymore.
%The phase diagram of QC$_2$D has
%been depicted based on the lattice
%simulation~\cite{Hands:2006ve,Hands:2010gd,Cotter:2012mb,Boz:2013rca}.
Through lattice simulations, actually, 
%In particular, 
the CSC phase has been found in
Refs.~\cite{Cotter:2012mb,Boz:2013rca} %and then there are the CSC phase 
in the deconfined regime.
But the parameters used in this case lead a result that the pion mass
is rather heavier than the physical one, $m_\pi/m_\rho \sim 0.8$ where $m_\pi$ and
$m_\rho$ are the $\pi$ and $\rho$ meson masses.
Therefore, to conclude that our present model is correct or not,
we will need more data of the
lattice QC$_2$D with physical quarks
masses.

%%%%%%%%%%%%%%%%%%%%
Another point to be noticed is that we find the CSC phase when the charge $q$ becomes large, $q>1.5$. 
In Sec. \ref{sec:backreaction}, an example is shown for $q=2$, where
the CSC phases are found in both the deconfinement and confinement
phases with a definite critical line. % as shown in the section 3. 
However, since $q(=2/N_c)\geq 2$ means $N_c\leq 1$, the holographic approach is not useful in this case. 
So we cannot trust any result for large $q$.
The other example of CSC phase realization may be possible in the lower space dimension.
For 2+1 space dimension, % is reduced to 2+1. 
an analysis has been given in \cite{Nishi} for the
superconductor of R-charge (not for the baryon charge) though the equations of motion are similar. % It is known that
%the dimensional reduction is expected when an impurity is added to the system. Then
So it would be possible to study the {CSC phase} in such lower dimensional cases by using the holographic model used here.

As for the chiral symmetry, the phase diagram in $\mu$-$T$ plane has been examined by solving the profile of probe branes 
%with the baryon number chemical potential %of the baryon number
based on a top-down model in the deconfinement phase \cite{Hori}. 
And a chiral symmetry breaking phase
has been found in the small $\mu$ region. 
However, in the present model, it is impossible to find such a broken phase in both the deconfinement
and confinement phases when the scalar is set as a field dual to the quark anti-quark operator
in order to invesigate the chiral symmetry.
%phase in the both approaches, the probe approximation and the back-reacted case.  
This is easily understood. Since %the introduce
the scalar field %of the quark and anti-quark which 
has no baryon charge ($q=0$), then the effective mass cannot be changed from the given mass $m^2=-4$.
As a result, the trivial solution of the scalar field is the stable one. 
%So there appears no
%factor of the instabiltiy of the scalar as its effective mass, which is rather fixed as $m_{eff}^2=m^2=-4$. 
In other words,
the present bottom up model is not suitable for studying the chiral symmetry.
%This model should be used only for finding the CSC phase. }
{It is important to improve the model in this point, however, it is out of the present task.}

{Finally, we give a brief comment on} the flavor degrees of freedom %of quarks as well as the quark 
and the mass of quarks.
Although, in this paper, we have not considered them,
%when they are considered, %By taking them into account, 
we would have a variety of phases such as two flavor 
CSC 
\cite{Erdmenger:2007ap,Erdmenger:2008yj}
and color-flavor locking \cite{Chen:2009kx,Alford:2007xm}. 
And once this is accomplished, it might be possible to study an interesting issue in high density QCD.

\vspace{.3cm}
%%%%%%%%%%%%%%%%%%%%%%%%%%%%%%%%%%%%
\section*{Acknowledgments}
One of the authors (M.T.) is supported by JSPS Grant-in-Aid for Scientific Research, Grant No.16K05357;
and another (K.K.) is supported by the Grants-in-Aid for Scientific Research,
Grant No. 18K03618.

%%%%%%%%%%%%%%%%%%%%%%%%%%%%%%%%%%%%%%

%%%%%%%%%%%%%%  References %%%%%%%%%%%%%%
 
\newpage
\end{document}